\documentclass[twocolumn,aps,showpacs,preprintnumbers,amsmath,amssymb]{revtex4}
\usepackage{graphicx}
\usepackage{dcolumn}
\usepackage{bm}

\begin{document}
\linespread{1.6}

\title{Resemblances and differences in mechanisms of noise-induced resonance} 

\author{R.~Centurelli$^{1}$, 
        S.~Musacchio$^{1,2}$,
        R.A.~Pasmanter$^{3}$ and     
        A.~Vulpiani$^{1,2,4}$}

\affiliation{$^{1}$ Department of Physics, 
University La Sapienza, P.le A. Moro 2, I-00185 Roma, Italy.  }
\affiliation{$^{2}$ INFM (UdR and CSM) - Unit\`a di Roma La Sapienza.} 
\affiliation{$^{3}$ K.N.M.I., P.O.Box 307, 3730 AE, De Bilt, The Netherlands}
\affiliation{$^{4}$ INFN - Sezione di Roma La Sapienza.}

\date{\today}

\begin{abstract}

Systems showing stochastic resonance (SR) 
or coherent resonance (CR) share some features, 
in particular the nearby periodic character of the signal. 
We show that in spite of this resemblance the different underlying 
dynamics can be detected in experimental data
by studying the histogram of inter-spikes
times and some statistical properties like two-times correlation 
functions.
We discuss the possible relevance for climate modeling. 

\end{abstract}


\maketitle

%%%%%%%%%%%%%%%%%%%%%%%%%%%%%%%%%%%%%%%%%%%%%%%%%%%%%%%%%%%%%%%%%%
\section{Introduction} 

The mechanism of stochastic resonance (SR) was initially introduced as a possible
explanation of climate changes on long time-scales~\cite{1}.
During the last two decades it has been applied to a wide class of systems
ranging from analog circuits, neurobiology, ring lasers, systems
with colored noise, etc; for a review see Ref.~\cite{2}.

The prototypical system showing SR, which is also the original one used to
model climate changes, is the stochastic differential
equation
\begin{equation}
  \frac{dx}{dt}=
 -\frac{\partial V(x,t)}{\partial x}
 +\sqrt{2 D}\eta \;,
\label{eq:1.1}
\end{equation}
where $\eta$ is a Gaussian, white noise with
$\langle \eta(t) \rangle = 0 $ and  
 $\langle \eta(t) \eta (t') \rangle = \delta (t - t')$,
$D$ measures the noise intensity
and $V(x,t)$ a is double well potential with a time periodic term
\begin{equation}
V(x,t) = \frac{x^4}{4} -\frac{x^2}{2} + A x \cos(2\pi t/T) \,\, .
\label{eq:1.2}
\end{equation}
In the case of a stationary potential, i.e., $A=0$, the jumps between the
two minima at $x=-1$ and $x=1$ are independent events whose probability
distribution is approximately Poissonian~\cite{R1}.
Using simple arguments based on the Kramers exit-time formula~\cite{3},
it can be shown that there is range of values of $D, \, T$
and $A$ where SR is present, i.e., the jumps between 
the two minima (close to $-1$ and $+1$ if $A$ is sufficiently small)
are strongly synchronized with the forcing
and that 
the probability distribution function (PDF) of the jumping 
time $\tau$ has a relatively sharp
peak around $T$ ~\cite{1,2}.

The phenomenon of SR provides one example of the nontrivial
role that noise can play in dynamical systems with an external periodic 
forcing.
Besides SR, there exist other 
examples of the ``constructive role'' of noise, e.g., 
one can have a synchronization of trajectories generated 
by different initial conditions and 
the same noise realization~\cite{R2}. Our interest will focus on cases 
where noise can enhance periodic behavior, e.g., the so-called
coherent resonance (CR) and the noise-induced dynamics in systems
with time delay (ND).

The phenomenon of CR~\cite{4}  has been found in models describing excitable
systems that occur in different fields like chemical reactions, neuronal
and other biological processes~\cite{5,5b}. 
The prototypical stochastic differential
equation used in this case is the FitzHugh-Nagumo system defined by:
\begin{eqnarray}
 \epsilon \frac{dx}{dt} & = &  x-\frac{x^3}{3} -y 
 \label{eq:1.3} \\
 \frac{dy}{dt}  & = &  x+a+\sqrt{2D}\eta \,\, 
\label{eq:1.4}
\end{eqnarray}
with $\epsilon \ll 1$ so that the time evolution of $x$ is much faster
than that of $y.$
For $|a|>1$ there is a stable fixed point, for $|a|<1$ 
there is an unstable fixed point and a limit cycle.
The cycle consists of two pieces of slow motion connected
by a fast jump.
If $|a|$ is slightly larger than $1$ the system is excitable~\cite{4} , i.e.,
small deviations from the fixed point may generate large pulses (also
called spikes)~\cite{6}.
Moreover, in this case, one finds that there is a range of values
of the noise intensity $D$ such that CR appears, i.e., roughly periodic
noise-excited oscillations are present, resembling the SR oscillations~\cite{4}.

The prototypical example for ND~\cite{7} is the over-damped particle motion
in the double-well potential $V(x(t),x(t-T))$:
\begin{eqnarray}
  { \frac{dx(t)}{dt} } & = & 
 -{ \frac{\partial V(x(t),x(t-T))}{\partial x(t)} }
 +\sqrt{2 D}\eta \nonumber \\
 & = & x(t) - x(t)^3 - A x(t-T) +\sqrt{2D}\eta
\label{eq:1.5}
\end{eqnarray}
where $T$ is the delay.
It is not very difficult to realize 
that the delay term $Ax(t-T)$ has a role similar to that of
the periodic forcing in Eq.~(\ref{eq:1.1} , \ref{eq:1.2}).
Accordingly, in a certain range of parameters' values , there is a
sort of periodic motion
with period $T$ or $2T$ (this depends upon the sign of $A$).
This ND equation had been proposed as a model for 
some climate changes~\cite{R3}. 

%%%%%%%%%%%%%%%%%%%%%%%%
\begin{figure}[t]
\begin{center}
\includegraphics[scale=0.7]{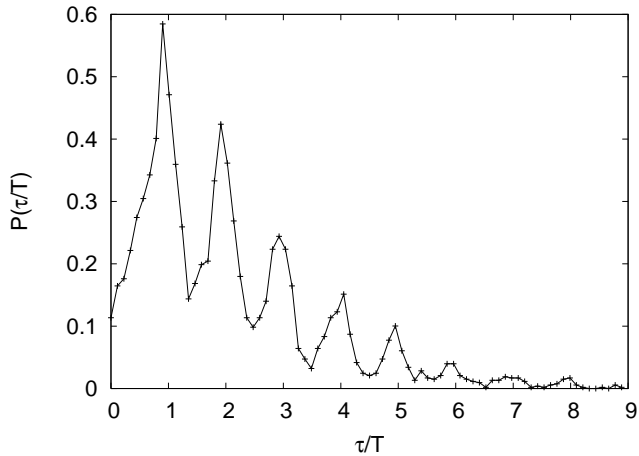}
\caption{PDF of the inter-spikes time 
in the system described by Eq.~(\ref{eq:1.1}) 
showing stochastic resonance.
The external periodic force has period $T=100$
and amplitude $A = -0.15$. The noise intensity is at the
corresponding optimal value $D^* = 0.10$. 
}
\label{fig:PdfSR}
\end{center}
\end{figure}
%%%%%%%%%%%%%%%%%%%%%%%%
%%%%%%%%%%%%%%%%%%%%%%%%
\begin{figure}[t]
\begin{center}
\includegraphics[scale=0.7]{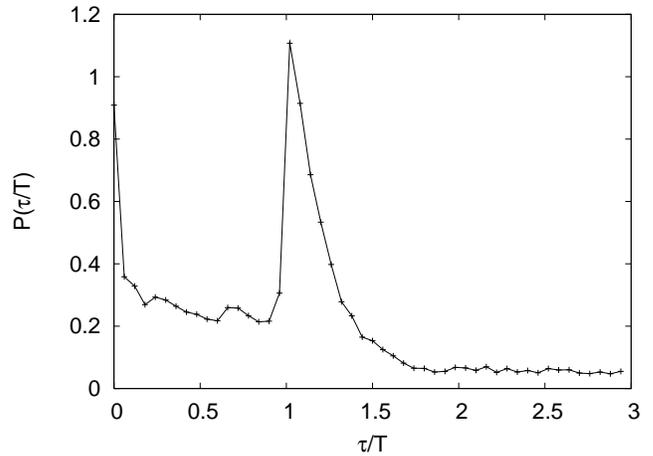}
\caption{PDF of the inter-spikes time 
in the system described by Eq.~(\ref{eq:1.5}).
The amplitude of the delay term is $A = - 0.15$, 
and the delay time is $T = 100$. 
The noise intensity is at the corresponding 
optimal value $D^* = 0.10$. 
}
\label{fig:PdfND}
\end{center}
\end{figure}
%%%%%%%%%%%%%%%%%%%%%%%%

Although SR, CR and ND, are similar phenomena,
in the sense that their time evolution is nearly periodic,
there are also some important differences. For example:
a)
due to the presence of the term $A x(t-T),$ Eq.~(\ref{eq:1.5})  is
in fact an infinite dimensional system since in order to determine
$x(t)$ for $t>0$ one has to specify $x(t')$ with $-T \le t' \le 0$.
On the contrary for Eq.~(\ref{eq:1.1}) it is sufficient to know $x(0)$.
b)
in the case of SR the periodicity is due to the external forcing
while in the CR case the periodicity has an internal origin, i.e.,
the periodic motion is due to the intrinsic dynamics
and, at variance with SR, its period cannot be changed
by tuning external control parameters. This difference can play an
important role in, e.g., the context of climate changes and glaciation,
for more details, refer to the last Section.

The aim of this paper is to analyse the differences among
SR, ND and CR and their possible relevance to applications.
In Sect II we will briefly review some properties of the PDF 
of the inter-spikes times for SR, ND and CR.
In particular, we recover a recent result~\cite{8} showing that
CR and SR are not conflicting or excluding mechanisms, i.e.,
the same  periodically driven system, 
e.g., the one given by Eqs.~(\ref{eq:1.3},\ref{eq:1.4})
 with a time-dependent parameter
$a(t)  = a_0 + a_1 cos(2 \pi t /T)$ can present a transition
from SR to CR behavior when the noise intensity $D$ is increased .
In Sect III we show that, in spite of some resemblance, SR, ND and
CR exhibit different statistical features which, at least in principle,
can be detected in experimental data. In particular we have that 
for SR the correlation function $C(\tau)$, after a transient period,
is periodic and does not relax to zero. 
On the contrary, for the CR and ND cases $C(\tau)$ shows damped 
oscillations.
Sect IV is devoted to general remarks and conclusions. In
particular we deal with the potential relevance of the differences
between SR, CR and ND to climate modeling.

%%%%%%%%%%%%%%%%%%%%%%%%%%%%%%%%%%%%%%%%%%%%%%%%%%%%%%%%%%%%%%%%%%%%%%%%%%%%%%
\section{Statistics of inter-spikes times}
\label{secII}

One basic feature shared by the three models introduced in the
previous Section,  
is the presence of two characteristic states: two equilibria in the case
of SR and ND, a rest state and an excited one in the FitzHugh-Nagumo model. 
Jumps between these states are made possible by the noise.
Moreover, it turns out that there exists an optimal value of the noise
intensity such that this jumping becomes approximately periodic, 
i.e., the typical time between two consecutive transitions is 
roughly constant. 

According to the terminology of biological systems, where CR 
was originally introduced, we will refer to the time interval between 
consecutive transitions as the ``inter-spikes time $\tau$''. 
In the case of the system~(\ref{eq:1.3},\ref{eq:1.4}) where 
the dynamical variables show well defined maxima, 
the definition of the inter-spikes time
 $\tau$ is rather natural. 
In the case of the other two systems~(\ref{eq:1.1},\ref{eq:1.2}) 
and~(\ref{eq:1.5}) one can define 
$\tau$ as $ \tau =  t_{n+1} -t_n$ where $t_n$ is the $n-$th crossing time, 
i.e.,
$x(t_n) = 0$ and $ \dot{x} (t_n) > 0 $.  
A measure of the signal's periodic character is provided by 
the normalized variance $NV$ of inter-spikes times,  
$NV = \sqrt{Var(\tau)} / \langle \tau \rangle $.
For generic noise intensity the transitions occur at random times, 
and $p(\tau)$, the PDF of inter-spikes times, is weakly localized, 
i.e., $NV$ is of order 1.  
We define the resonant or optimal value of the noise intensity, and
denote it by $D^*$, as the value of $D$
for which the system has minimal normalized variance $NV$.

%%%%%%%%%%%%%%   Fig. (3)     %%%%%%%%%%%%%%%%%%%%%
\begin{figure}
\begin{center}
\includegraphics[scale=0.7]{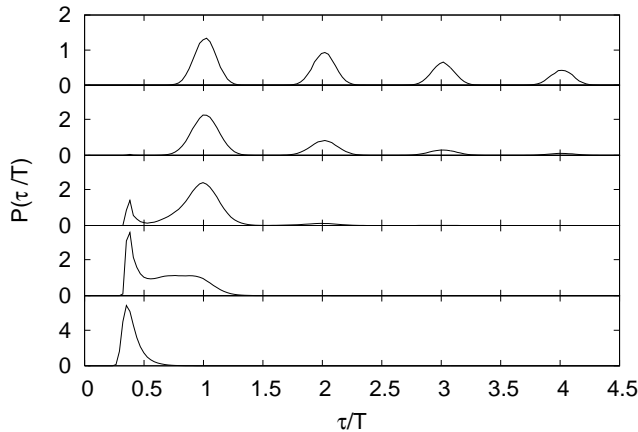}
\caption{
PDF of the inter-spikes time in the
FitzHugh-Nagumo model Eqs.~(\ref{eq:1.3},\ref{eq:1.4})
with parameters' values
$a_0 = 1.05$,
$a_1 = 0.04$,
$\epsilon = 0.01$
and
$T=10 > T_*=3.7$
for different noise intensities. From top to bottom,
$D = 2.5 \times 10^{-5}$,
$D = 4.3 \times 10^{-5}$,
$D = 1.1 \times 10^{-4}$,
$D = 2.3 \times 10^{-4}$
and
$D = 4.3 \times 10^{-3}$. 
}
\label{fig:PdfFN}
\end{center}
\end{figure}
%%%%%%%%%%%%%%%%%%%%%%%%

%%%%%%%%%%%%%%%%%%%%%%%%
\begin{figure}
\begin{center}
\includegraphics[scale=0.7]{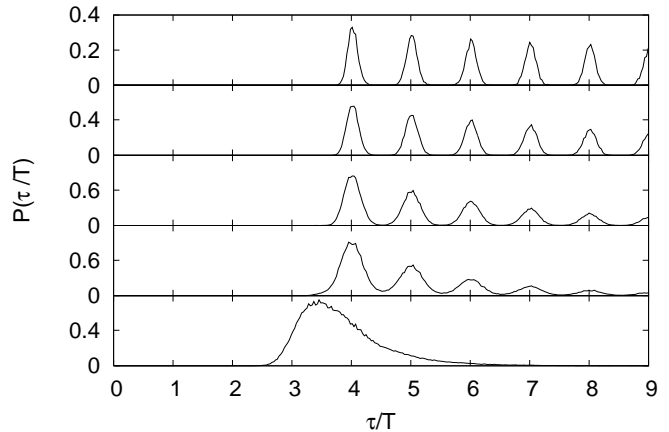}
\caption{
As in Fig.~(\ref{fig:PdfFN} )
but with 
$T=1 < T_*=3.7$.    
}
\label{fig:isto1}
\end{center}
\end{figure}
%%%%%%%%%%%%%%%%%%%%%%%%

In Figure~{\ref{fig:PdfSR}} we show the 
PDF of the inter-spikes time interval in the case of SR
for the system described by Eq.~(\ref{eq:1.1})
at the optimal noise intensity $D^*=0.1$. We see that $p(\tau)$ 
is peaked around $\tau = nT$ with 
$ n=1,2,.....  $, 
and that the envelope is approximately exponential.
This feature can be easily explained as follows. 
Consider the trajectories $x(t)$, starting at $t=0$ 
from the favored well, i.e. $x(0)= 1$ if $A < 0$. 
In the case of SR, at $t$ close to $T/4$ many of the trajectories 
will jump onto the other minimum at $x= -1$ and after half period they 
will jump back again onto $x = 1$. However, a fraction of the trajectories
remains in the ``wrong position'' (i.e. in the unfavored well)
for $t$ close to $T$.
Calling $P$ the probability of this event, we have that  
the integral of $p(\tau)$ around $T$, say for $\tau \in [ 0.5 T , 1.5 T ]$, is $(1-P)$.
The events with inter-spikes time $\tau \sim nT$ 
 correspond to trajectories $x(t)$ 
which are in the the ``wrong'' minimum at $T, 2T, ....,nT$.
Taking into account the periodicity of the forcing and assuming that
the system's memory is much shorter than the external period $T$
we have that the probability to have $x(t)$ in the ``wrong'' minimum
at  $t \sim kT$ under the condition that $x(t)$ was in  
the the ``wrong'' minimum
at  $t \sim (k-1)T$, does not depend on the behavior for $t<(k-1)T$.
Therefore  the  integral of $p(\tau)$ around $nT$ is 
$(1-P)P^{n-1} \sim e^ {-c n}$ with $c=- \ln P$. The envelop
of the inter-spikes-interval histogram has recently 
been computed by Berglund and Gentz~\cite{BG}
in a more general and rigorous setting.

In the case of CR (not shown) and ND, see Figure~\ref{fig:PdfND},
 the PDF of the inter-spikes 
time interval is peaked around $T$ where $T$ is the characteristic internal 
time of the system, namely the delay time in the ND case, 
and the period of the limit 
cycle in the CR case. 
The parameters of the memory term in the
ND-model~(\ref{eq:1.5}) have been
chosen in order to emphasize the similarities with the SR-model,
i.e., the amplitude $A$ 
and delay time $T$ are identical to the amplitude 
and period of the external forcing in Eq.~(\ref{eq:1.1}).
One consequence of this choice is that the intensity of the optimal noise coincides with that in the SR-model. 
Notice that, 
at variance with the SR case, no peaks are present at multiples of $T$.

In summary, in all three systems there 
exists an optimal noise intensity which 
%``regularize'' their behavior, i.e.
produces a roughly periodic signal $x(t)$. 
At this intensity a sharp peak appears in the 
inter-spikes time PDF. 
In the case of CR and ND the PDF has only one maximum at 
$\tau = T$. On the other hand, in the case of SR 
other maxima appear at $nT$. This effect can be considered 
as one of the distinctive marks of the stochastic resonance.
In contraposition to the CR and ND cases, 
the periodicity of the signal in the SR case is 
induced by an external periodic force, which triggers 
the jumps at fixed times $t=nT$ in such a way that 
the system synchronizes with this external ``clock".  

%%%%%%%%%%%%%%%%%%%%%%%%
\begin{figure}[t!]
\begin{center}
\includegraphics[scale=0.7,draft=false]{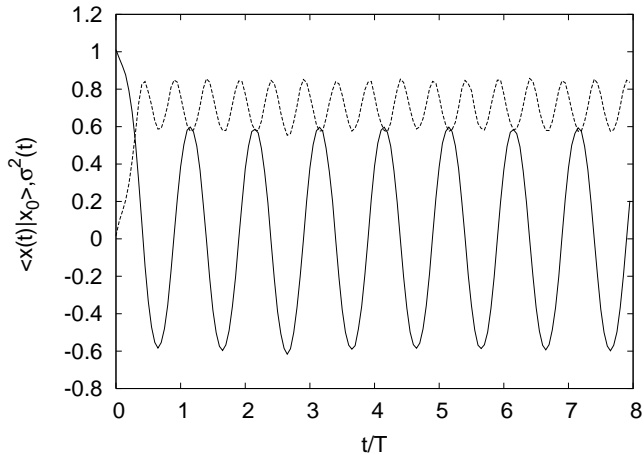}
\caption{Conditional average 
$\langle x(t)|x_0 \rangle$
(solid line)
and conditional variance
$\sigma^2(t) $
(dashed line) 
in the case of SR for the system described by Eq.~(\ref{eq:1.1}). 
The initial position is $x_0 = 1$, i.e., in the favored well.  
The external force has period $T=100$
and amplitude $A = -0.15$. The noise intensity is at its 
optimal level $D = 0.10$.
The average is taken over $N=10^4$ realizations of the noise.  
} 
\label{fig:XSR}
\end{center}
\end{figure}
%%%%%%%%%%%%%%%%%%%%%%%%
%%%%%%%%%%%%%%%%%%%%%%%%%%%%%%%%%%%%%%%%%%%%%%%%%%%%%%%
% \subsection{Coexistence of CR and SR}

The combined effects of noise and nonlinearity can result 
in more complex behavior~\cite{yacomotti,mendez}. 
In particular it has recently been shown in~\cite{8}
that SR and CR can coexist in the same system.
This behavior occurs, e.g., in the model~(\ref{eq:1.3}-\ref{eq:1.4}) 
when small oscillations are imposed on the control parameter 
$ a(t) = a_0 + a_1 \cos({2 \pi t /T})$. 
We consider only $a_0$ and oscillation amplitudes $a_1$
such  that the control parameter $a(t)$ never crosses the critical value, 
i.e. $ a(t) > 1, \forall t$.  
The presence of these small oscillations determines privileged 
times $t_n = (2n+1)T/2$  
at which the system is closer to the excited state, 
i.e., $a(t_n)$ approaches 1 from above,
and a noise induced transition is facilitated. Analogously to the case 
of the double well system with periodic forcing, 
there exists an optimal noise intensity for which a regular,
quasi-periodic behavior emerges. More precisely:
at low noise intensity, $p(\tau)$ 
has rather sharp peaks at $\tau = nT$ with an 
approximately exponential envelope, 
showing all the features of SR, 
see the two upper panels in Figure~{\ref{fig:PdfFN}}.
At higher noise intensity, as in the two lower panels in
Figure~{\ref{fig:PdfFN}}, the small oscillations
in $a(t)$ become irrelevant and the system shows CR 
as if the control parameter were fixed at its mean value $a_0$.
The shape of $p(\tau)$ behaves accordingly
as the noise increases: the maxima of SR diminish and a  
single peak with an exponential tail appears at $\tau = T_*$ where $T_*$ is the
period of the system's limit cycle. 

Moreover, when the external 
force period $T$ is shorter than the internal, limit-cycle period $T_*$ 
there are peaks only for $nT \ge T_*$ because once the excited state 
is reached the system needs at least a time
$T_*$ in order to relax back into the excitable, rest state and restart
the whole cycle (see fig.~\ref{fig:isto1}). 

%%%%%%%%%%%%%%%%%%%%%%%%
\begin{figure}[t!]
\begin{center}
\includegraphics[scale=0.7,draft=false]{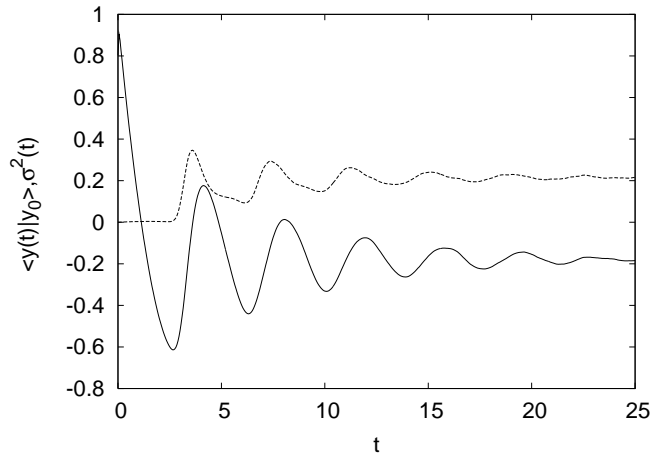}
\caption{Same as in Fig.~\ref{fig:XSR}
in the case of CR for the system described by 
Eqs.~(\ref{eq:1.3},\ref{eq:1.4}) 
with $a=1.05$, $\epsilon = 0.01$ and 
optimal noise intensity $D = 2.5 \times 10^{-3}$. 
The initial position is $x_0 = 2, y_0 = 0.8$, i.e., in the 
excited state.  
The average is taken over $N=10^4$ realizations of the noise.
}  
\label{fig:XCR}
\end{center}
\end{figure}
%%%%%%%%%%%%%%%%%%%%%%%%

%%%%%%%%%%%%%%%%%%%%%%%%%%%%%%%%%%%%%%%%%%%%%%%%%%%%%%%%%%%%%%%%%%%%
\section{Conditional averages and correlation functions}

If one would observe just a single trajectory, SR, CR and ND
would appear rather similar since the three cases would
present us with a
nearly periodic $x(t)$.
An analysis based on Fourier spectra, as it is often done,
would reinforce this picture.
In the previous Section we contrasted the multi-peaked PDF
of inter-spikes time in the SR case with the one-peaked PDF
in the CR and ND cases. 
In this Section we bring to the fore some statistical properties
which are present in the SR case but are absent both in the CR and 
in the ND cases.

Consider an ensemble of $N$ trajectories
$\{ x^{(n)}(t), n=1,...., N \gg 1 \},$  sharing the same initial conditions
$x^{(n)}(0)= x_0$, but with different realizations of the noise
$\eta (t)$
and compute from Eqs.~(\ref{eq:1.1}),~(\ref{eq:1.3}-\ref{eq:1.4}) 
and~(\ref{eq:1.5}) 
the conditional average
$\langle x(t)|x_0 \rangle$, 
\begin{equation}
\langle x(t)|x_0 \rangle = \frac{1}{N} \sum_{n=1}^N x^{(n)}(t) \;,
\label{eq:3.1}
\end{equation}
and the conditional variance
\begin{eqnarray}
\sigma^2(t) & = & \langle x^2(t)|x_0 \rangle - \langle x(t)|x_0 \rangle ^2 \nonumber \\
& = & \frac{1}{N} \sum_{n=1}^N [x^{(n)}(t) - \langle x(t)|x_0 \rangle ]^2 \;,
\label{eq:3.2}
\end{eqnarray}

%%%%%%%%%%%%%%%%%%%%%%%%
\begin{figure}[t]
\begin{center}
\includegraphics[scale=0.7]{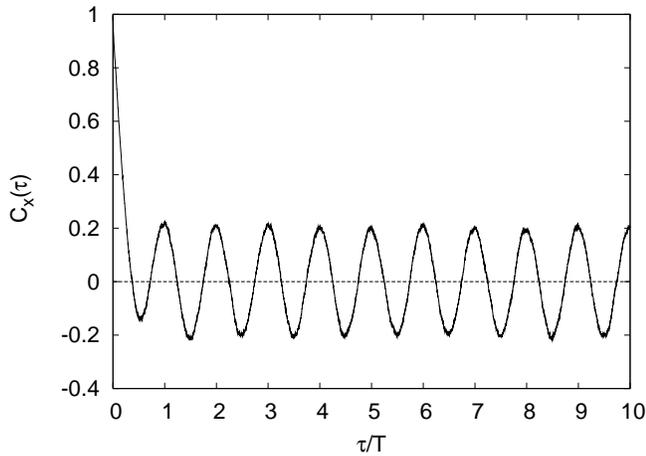}
\caption{The periodic correlation function $C(\tau)$ 
of the SR model Eq.~(\ref{eq:1.1}).
Parameters' values as in Fig.~\ref{fig:PdfSR}
} 
\label{fig:CorrSR}
\end{center}
\end{figure}
%%%%%%%%%%%%%%%%%%%%%%%%

In Figs.~\ref{fig:XSR} and \ref{fig:XCR} we show 
$ \langle x(t)|x_0 \rangle $ and $\sigma^2(t)$
as functions of the time $t$, for the SR case and the CR case
respectively. In both cases the noise-intensity 
values $D$ are the optimal ones. 
In the SR case $\langle  x(t)|x_0 \rangle $ does 
not relax to zero at large times $t$
and, ignoring the initial transient, it is periodic.
The conditional variance $\sigma^2(t)$ reaches its minima
when the absolute value of the  conditional  average 
$ \langle x(t)|x_0 \rangle $
reaches its maxima. The largest values for $\sigma^2(t)$ 
are achieved
when  $\langle x(t)|x_0 \rangle $ is around zero.
In other words, the main uncertainty in the process occurs
around $T/4, 3T/4, 5T/4 $ and so on, i.e., when the jumps between
the two minima take place.

As it can be seen in Fig~\ref{fig:XCR} the behavior found in the CR
case is different.
Even with the noise intensity at its optimal value, 
after a few damped oscillations both the conditional average and variance 
relax to the constant values 
$ \langle \langle x(t) \rangle \rangle $ and 
$\langle \langle x^2(t) \rangle \rangle  
- \langle \langle  x(t)  \rangle \rangle^2$, 
where $\langle \langle \cdots \rangle \rangle $ indicates a 
time average. 

This behavior underlines the intrinsic difference between the SR and 
CR mechanisms. 
In the case of SR, the presence of an external synchronizing
 force, makes the 
transitions from state $+1$ to state $-1$ to occur around 
preferred times. 
A set of independent replicas, 
initially localized in one of the two wells, will therefore 
quickly reach a periodic configuration, with the 
maximum probability localized in the time-dependent favored well.
On the contrary, in the CR case, there are no externally defined
preferred times for the transitions, 
therefore each replica quickly loses its initial 
synchronization with the other ones and after a few periods the 
jumps occur at different times for different replicas. 

\begin{figure}[t!]
\begin{center}
\includegraphics[scale=0.7]{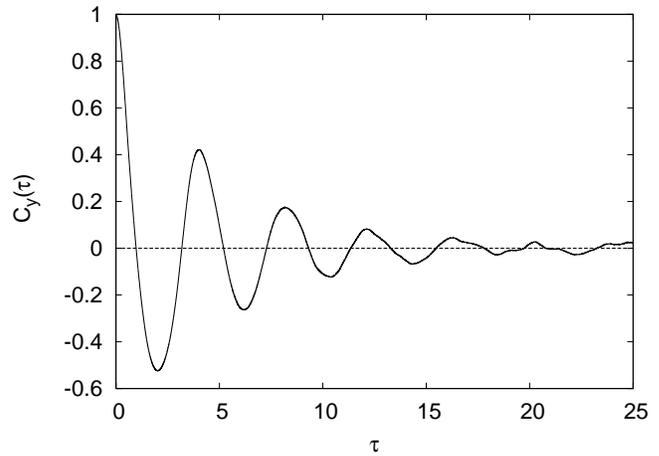}
\caption{Correlation function $C(\tau)$ 
of the $y$ variable for the FitzHugh-Nagumo 
model described 
by Eqs.~(\ref{eq:1.3} , \ref{eq:1.4}).
Parameters' values as in Fig.~\ref{fig:XCR}
}
\label{fig:CorrCR}
\end{center}
\end{figure}

The correlation function
\begin{equation}
C(\tau)= 
{ 
\frac 
{\langle \langle x(t+\tau)x(t) \rangle \rangle 
- \langle \langle x \rangle \rangle^2} 
{\langle \langle x^2 \rangle \rangle - 
\langle \langle x \rangle \rangle ^2}
},
\label{eq:3.3}
\end{equation}
behaves similarly to
$\langle x(t)|x_0 \rangle$: in the CR case it relaxes to zero while 
in the SR case it remains periodic with non decreasing amplitude, 
see Figs.~(\ref{fig:CorrSR}) and (\ref{fig:CorrCR}).
If one defines a correlation time $\tau_c$ as
\begin{equation}
\tau_c = \int_0^{\infty} C(\tau)^2 d \tau ,
\label{eq:3.4}
\end{equation}
one finds that $\tau_c$
diverges in the SR case while it remains finite in the CR 
case and that,
as a function of the noise intensity, 
it attains its maximum $\tau_c = 1.03$
at the optimal noise intensity value.
The behavior in the ND case (not shown) is 
qualitatively very similar to that obtained in the CR case. 

%%%%%%%%%%%%%%%%%%%%%%%%
%%%%%%%%%%%%%%%%%%%%%%%%

%%%%%%%%%%%%%%%%%%%%%%%%%%%%%%%%%%%%%%%%%%%%%%%%%%%%%%%%%%%%%%%%%%%%%%%%%%%%
\section{Conclusions and Discussion}

The numerical study of SR, CR and ND presented in this paper
is focused on contrasting statistical features
which would be difficult or impossible to detect by simply
looking at the spectrum since
the spectrum shows a peak that achieves
a maximal sharpness for the optimally chosen noise intensities
in all the cases we have studied.
As we have seen, it is possible to distinguish between the
different underlaying dynamical mechanisms by studying 
the PDF of inter-spikes times, conditional 
averages and time-delayed correlation functions.

We believe that these results are of interest not only in the
context of dynamical systems but also in the study of certain
climate phenomena.
For time scales of order $O(10^5 ys)$ and larger,
Milankovich\cite{Milan}
has proposed that Earth's climate has been determined by the
influx of solar energy to such an extent that the fluctuations
in, e.g., the global mean temperature and seasonality 
must have been closely
correlated with the variations in the incoming energy flux due
to the periodicities in Earth's orbit.

Another possible issue, for which the presented results are
potentially interesting, is
the so-called Dansgaard-Oeschger(DO) events~\cite{DO}
which have been inferred from the study of
Late-Pleistocene ice cores and marine sediments.
These measurements show rapid warmings
of the atmosphere followed by a much
slower decay back into the average glacial conditions.
The warmings took place on a time scale of a few decades while
the relaxation back into glacial temperatures lasted some centuries
up to millennia.
They seem to have occurred at intervals of $1.500 \pm 200$ years
or integer multiples hereof~\cite{Schulz}.
They were absent during the Holocene, i.e., during the $10^4$ 
years before present.
The discovery of these rapid warmings led to proposing
a number of possible explanations,
some of them favoring the internal origin of the period
approximately equal to $1.500$ years~\cite{Broecker,Winton,Sakai},
while other explanations assume a similar period due to an external
astronomical forcing~\cite{Ganopolski}, i.e. a SR-type scenario.
In particular, it was shown that in an ocean-circulation box model 
and within an appropriate parameters' range,
the purely deterministic system has a fixed point and does not show
any time dependence while the addition of noise
leads to the generation of spikes with a well defined
inter-spikes time interval~\cite{Axel},
i.e., that coherence resonance is present in this ocean circulation
model.
As discussed in Section II, now we know that, at least
in some systems, it is possible to observe 
either CR or SR behavior depending upon the noise
intensity.

We have shown that it is
possible to distinguish between the SR and CR, e.g., by looking
at time-delayed correlation functions and at the PDF of the inter-spikes
times.
Needless to say, in order to compute such correlation functions,
or the PDF, a sufficiently long and accurate
series of measurements is required.
From the limited information about the
DO events that has been extracted from the geological record
it is difficult to decide in favor of one scenario or the other.
Indeed, the PDF for the inter-spikes time is qualitatively
in agreement with the one observed in the SR case.
On the other hand  the time values at which one observes
peaks of the PDF do not correspond to known astronomical
periods, accordingly, a CR scenario would seem
more appropriate.

%%%%%%%%%%%%%%%%%%%%%%%%%%%%%%%%%%%%%%%%%%%%%%%%%%%%%%%%%%%%%%%%%%%%%%%%%%%%%%%%%%%%%
\section*{Acknowledgments}
We thank A.~Timmermann and A.S.~Pikovsky for stimulating discussions and useful comments.
This work has been supported by
MIUR-COFIN03 "Complex Systems and Many-Body Problems"
(2003020230)  and the  INFM (Statistical Mechanics 
and Complexity Center, Rome). 

%%%%%%%%%%%%%%%%%%%%%%%%%%%%%%%%%%%%%%%%%%%%%%%%%%%%%%%%%%%%%%%%%%%%%%%%%%%%%%%%%%%%%

%%%%%%%%%%%%%%%%%%%%%%%%%%%%%%%%%%%%%%%%%%%%%%%%%%%%%%%%%%%%%%%%%%%%%%%%%%%%%

\end{document}